\documentclass[prb,superscriptaddress,twocolumn,showpacs]{revtex4-1}
\usepackage[english]{babel}
\usepackage{amsmath,amssymb,amsfonts}
\usepackage{graphicx}
\usepackage{dcolumn}
\usepackage{bm}
\usepackage{bbm}
\usepackage{hhline}
\usepackage{braket}
\usepackage{placeins}

\begin{document}
\title{Anisotropic $g$ factor in InAs self-assembled quantum dots}

\author{Robert Zielke}
  \affiliation{Department of Physics, University of Basel,
    Klingelbergstrasse 82, 4056 Basel, Switzerland}
\author{Franziska Maier}
  \affiliation{Department of Physics, University of Basel,
    Klingelbergstrasse 82, 4056 Basel, Switzerland}
\author{Daniel Loss}
  \affiliation{Department of Physics, University of Basel,
    Klingelbergstrasse 82, 4056 Basel, Switzerland}
  \affiliation{CEMS, RIKEN, Wako, Saitama 351-0198, Japan}

\begin{abstract}
We investigate the wave functions, spectrum, and $g$-factor anisotropy of low-energy electrons confined to self-assembled, pyramidal InAs quantum dots (QDs) subject to external magnetic and electric fields. 
We present the construction of trial wave functions for a pyramidal geometry with hard-wall confinement. 
We explicitly find the ground and first excited states and show the associated probability distributions and energies. 
Subsequently, we use these wave functions and 8-band $\bm{k\cdot p}$ theory to derive a Hamiltonian describing the QD states close to the valence band edge. 
Using a perturbative approach, we find an effective conduction band Hamiltonian describing low-energy electronic states in the QD. 
From this, we further extract the magnetic field dependent eigenenergies and associated $g$ factors.
We examine the $g$ factors regarding anisotropy and behavior under small electric fields. 
In particular, we find strong anisotropies, with the specific shape depending strongly on the considered QD level. Our results are in good agreement with recent measurements [Takahashi {\textit{ et al.}}, Phys. Rev. B \textbf{87}, 161302 (2013)] and support the possibility to control a spin qubit by means of $g$-tensor modulation.
\end{abstract}


\maketitle

\section{Introduction}
Electron spins confined to semiconductor quantum dots (QDs) are excellent candidates for the physical realization of qubits, the elementary units of quantum computation.\cite{loss_quantum_1998}
The qubit state can be initialized and manipulated by means of externally applied electric and magnetic fields.
Thus knowledge about the qubit's response to these fields is crucial for the successful operation of qubits.
This response depends strongly on the type of QD  considered,  e.g.,\  lateral gate defined QDs, nanowire QDs, and self-assembled QDs.\cite{kloeffel_prospects_2013,hanson_spins_2007}
The most prominent type of QDs for self-assembled QDs are InAs QDs grown on a GaAs surface or in a GaAs matrix. 
These QDs can be grown in various shapes such as pyramids,\cite{grundmann_inas/gaas_1995,grundmann_ultranarrow_1995,ruvimov_structural_1995} truncated pyramids,\cite{ledentsov_direct_1996} and flat disks\cite{bayer_coupling_2001} and hence are highly strained due to the lattice constant mismatch of substrate and QD materials.
In self-assembled InAs QDs, spin states have been prepared with more than 99\% fidelity\cite{atature_quantum-dot_2006} and complete quantum control by optical means has been shown.\cite{press_ultrafast_2010,press_complete_2008} 
However, full qubit control by means of external fields and small system sizes are the most important goals in solid state  based quantum computation, allowing for the construction of integrated circuits.\cite{kloeffel_prospects_2013,hanson_spins_2007}
Regarding these requirements, $g$-tensor modulation is a powerful mechanism that allows control of the qubit\cite{pingenot_electric-field_2011,pingenot_method_2008,bennett_voltage_2013} but is sensitive to the shape of the QD hosting the qubit.\cite{hanson_spins_2007}
Hence the qubit behavior under the influence of geometry, external fields, etc.,\ is still subject to ongoing scientific effort.\cite{troiani_hyperfine-induced_2012,petersson_circuit_2012,jin_strong_2012} 
A crucial ingredient  for modeling the qubit behavior is the knowledge of the particle distribution within the QD, i.e.,\ the envelope wave function which is mainly determined by the shape of the QD. 
For simple structures such as spheres, flat cylinders, and cubes, the wave functions in QDs can be described analytically, e.g.,\ by employing hard-wall or harmonic confinement potentials.\cite{cohen-tannoudji_quantum_1977} 
For more complicated shapes usually numerical models are employed.\cite{grundmann_inas/gaas_1995, pryor_electronic_1997,pryor_eight-band_1998,stier_electronic_1999,de_calculation_2007} 
Recently, there have been efforts to find analytical wave functions for pyramids with different types of boundary conditions.\cite{horley_particle_2012,vorobiev_effect_2013} 
However, the set of wave functions introduced so far has been observed to be incomplete, lacking for example the ground state wave function. 
Both analytical and numerical methods are employed to further explore QD characteristics such as strain,\cite{grundmann_inas/gaas_1995,nenashev_strain_2010} spectra,\cite{grundmann_inas/gaas_1995} and $g$ factors.\cite{pryor_lande_2006,*pryor_erratum:_2007}
Explicit values depend on the material properties.
Building QDs in materials with very large, isotropic bulk $g$ factors, i.e.,\ InAs ($g=-14.9$), is favorable due to an improved opportunity of $g$-factor modification.
Measurements emphasize the decrease of the $g$ factor when considering electrons in InAs QDs.
Numerical calculations\cite{de_calculation_2007,pryor_lande_2006,pryor_erratum:_2007} and measurements\cite{bayer_electron_1999,medeiros-ribeiro_spin_2002} show that the $g$ factor can go down to very small values and depends strongly on the dot size. 
Furthermore, recent measurements show a significant anisotropy\cite{dhollosy_g-factor_2013,takahashi_electrically_2013} of the $g$ factor which  turned out to be tunable by electrical means.\cite{takahashi_electrically_2013,Deacon_electrically_2011} 
This behavior of $g$ can be attributed to material- and confinement-induced couplings between the conduction band (CB) and the valence band (VB) which result in totally mixed low-energy states. 

The outline of this paper is as follows.
In Sec.~\ref{sec:model}, we present an 8-band $\bm{k\cdot p}$ Hamiltonian describing the low-energy QD states which accounts for strain and external electric and magnetic fields.
Additionally, we introduce a set of trial wave functions satisfying the hard-wall boundary conditions of a pyramidal QD.
Furthermore, we derive an effective Hamiltonian describing CB states in the QD.
In Sec.~\ref{sec:results}, we present the results of our calculations, in particular the $g$-factor anisotropy of CB QD levels.
These results are discussed and compared to recent measurements in Secs.~\ref{sec:discussion} and \ref{sec:comparison}, respectively.
Finally, in Sec.~\ref{sec:conclusion}, we conclude. 
\begin{figure}[tbh]
\centering
\includegraphics[width=.4\textwidth]{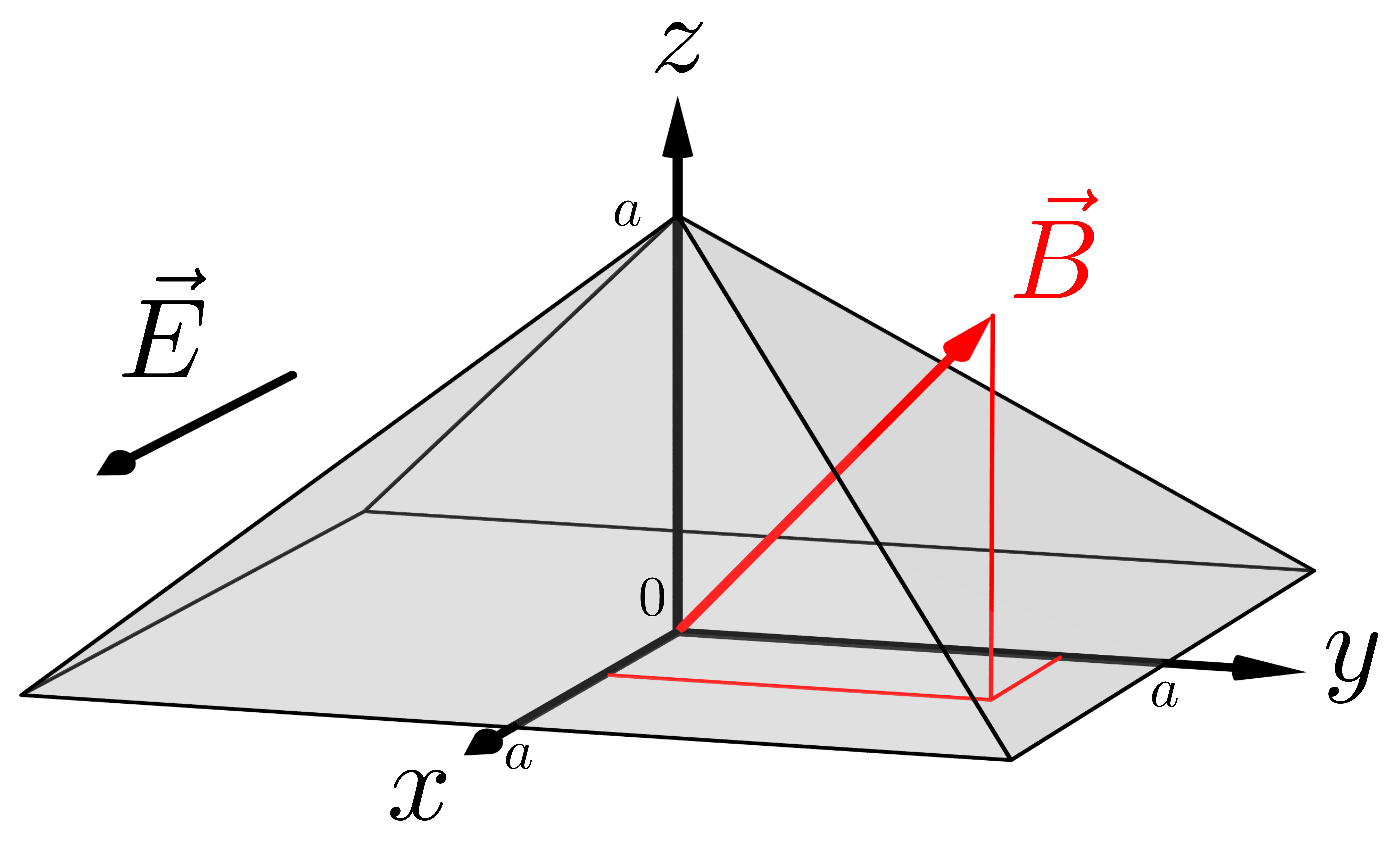}
\caption{Sketch of the QD geometry and the coordinate system used in this work with
$x$, $y$, and $z$ axes pointing along the growth directions [100], [010], and [001], respectively.
The externally applied fields under consideration are $\bm{B}=(B_x,B_y,B_z)$ and $\bm{E}=(E_x,0,0)$.}
\label{fig:setup}
\end{figure}

\section{Model \label{sec:model}}
In this section, we introduce the Hamiltonian and wave functions used in this work. 
Furthermore we outline the performed calculations and give the main results in a general manner. 
\subsection{Hamiltonian}
Low-energy states in bulk III-V semiconductors are well described by an 8-band $\bm{k\cdot p}$ model,\cite{winkler_spin-orbit_2003} which includes the CB and the VB consisting of heavy- (HH) and light-hole (LH) bands, and split-off (SO) bands.
The associated Hamiltonian $H_{\bm{k\cdot p}}$ is given in terms of two-fold degenerate  basis states $\ket{j,\pm}$, $j = \text{CB}, \text{HH}, \text{LH}, \text{SO}$, which are linear combinations of products of angular momentum eigenfunctions and real spin states.\cite{winkler_spin-orbit_2003} 
We model a pyramidal QD by taking into account a three-dimensional hard-wall confinement potential $V_c$ defining a square pyramid of height $a$ and base length $2a$ as sketched in Fig.~\ref{fig:setup}.
We introduce strain by adding the strain Hamiltonian $H_{\mbox{\tiny strain}}$.\cite{winkler_spin-orbit_2003} 
An analytical description of the strain distribution within an InAs pyramid enclosed in a GaAs matrix can be modeled by exploiting the analogy to electrostatic theory.\cite{nenashev_strain_2010}
We include the effect of an externally applied magnetic field $\bm{B}=\bm{\nabla\times A}$ defined by the vector potential $\bm{A}$ ($\bm{\nabla\cdot A}=0$) by adding two terms.
The first term is the magnetic interaction term $H_Z$.\cite{winkler_spin-orbit_2003} 
To derive the second term, $H_B$, we replace $\bm{k}\rightarrow\bm{k}+e\bm{A}/\hbar$ in $H_{\bm{k\cdot p}}$ and $H_{\mbox{\tiny strain}}$ in a semiclassical manner, where $e$ is the positive elementary charge and $\hbar$ the Planck constant. 
We drop all contributions independent of $\bm{B}$ and obtain a Hamiltonian which accounts for orbital effects of $\bm{B}$.
An external electric field $\bm{E}$ is included by adding the electric potential $H_E=e\ \bm{E\cdot r}$, with $\bm{r}=(x,y,z)$.
The full system is then described by the Hamiltonian
\begin{equation}
\label{eq:hamiltonian}
H=H_{\bm{k\cdot p}}
+H_{\mbox{\tiny strain}}
+H^{\mbox{\tiny }}_{Z}
+H^{\mbox{\tiny }}_{B}
+H^{\mbox{\tiny }}_{E}
+V_{c}.
\end{equation}
Note that literature values for $\bm{k\cdot p}$ parameters are usually given for 4-band models. In an 8-band model, the parameters have to be modified accordingly.\cite{winkler_spin-orbit_2003}

\subsection{Hard-wall wave function}
As a first step, we consider $V_{c}$ of a pyramidal QD analytically and require a vanishing particle density at the boundaries.
We construct a trial wave function satisfying these boundary conditions as follows. 

The Schr\"odinger equation of a particle confined in a square with sides of length $a$ with vanishing boundary conditions on the borders, has the well-known solution $\psi_{mn}^{\Box}(x,y)$ with eigenenergies $E^{\Box}_{mn}$. 
The wave function of a particle confined in an isosceles triangle obtained by cutting the square along the diagonal, $\psi^{\triangle}(x,y)$, is then constructed by linear combinations of degenerate solutions $\psi_{mn}^{\Box}(x,y)$ while requiring a vanishing wave function at the diagonal of the square.\cite{li_particle_1984}
We span the full three-dimensional (3D) volume of the pyramid and the corresponding wave functions with the product of two such triangles and the associated $\psi^{\triangle}$. 
This consideration suggests then the following ansatz  for the hard-wall wave functions inside the pyramidal geometry of the form
\begin{equation}
\begin{split}
\label{eq:3D-wave-function}
\psi_{\bm{m}}(\bm{r})
=&\ 
c\prod_{\xi=x,y}
\Big[
\sin\left(\alpha_{\xi}\ \xi^{+}\right)
\sin\left(\alpha_{z}\ \xi^{-}\right)
\\&
-(-1)^{m_{\xi}+m_{z}}
\sin\left(\alpha_{z}\ \xi^{+}\right)
\sin\left(\alpha_{\xi}\ \xi^{-}\right)
\Big],
\end{split}
\end{equation}
with
$\bm{r}=(x,y,z)$, $c=\csc(\pi z)/N_{\bm{m}}$, 
$\alpha_{i}=m_{i}\pi/a$, $m_{i}=1,2,3,\dots$, $m_{x}\ne m_{z}$, $m_{y}\ne m_{z}$, $\bm{m} = (m_x, m_y, m_z)$, 
$\xi^{\pm}=\xi\pm(z-a)/2$, and $N_{\bm{m}}$ such that the integral over the pyramid volume $\int d^3r\ |\psi_{\bm{m}}(\bm{r})|^2\equiv 1$.
We define energies of $\psi_{\bm{m}}(\bm{r})$ by taking
\begin{equation}
\frac{\hbar^2}{2 m_0}\bra{\psi_{\bm{m}}(\bm{r})}(-i \nabla)^2 \ket{\psi_{\bm{m}}(\bm{r})} = E_{\bm{m}}, 
\label{eq:HpsiEpsi}
\end{equation}
where $m_0$ denotes the bare electron mass.
For notational simplicity we use $\psi_{\bm{m}}\equiv\psi_{m_xm_ym_z}$ and $E_{\bm{m}}\equiv E_{m_xm_ym_z}$.
Exact analytical solutions of the Schr\"odinger equation have been derived using specular reflections of plain waves at the boundaries of the geometry.\cite{horley_particle_2012}
However, the obtained set of solutions is incomplete, consisting solely of excited states and especially lacking the ground state. 
We stress that our ansatz $\psi_{\bm{m}}$ is not an eigenstate of the Schr\"odinger equation. 
However, the energies $E_{\bm{m}}$ we find are lower than the eigenenergies of the Schr\"odinger equation derived in Ref.~\onlinecite{horley_particle_2012}; see Secs.~\ref{sec:res:prob} and \ref{sec:disc:prob}.
In addition, the  wave function for the lowest energy state, $\psi_{221}$, exhibits the expected nodeless shape for the ground state.
A more detailed justification of $\psi_{\bm{m}}(\bm{r})$ is given in Appendix \ref{app:wave-function}.
In the following calculations, we apply these trial envelope wave functions for both CB and VB states. 
In general, electron and hole envelope wave functions differ;\cite{grundmann_inas/gaas_1995,stier_electronic_1999} however, this choice is justified since we find that even this overly simplified picture yields already good results.

\subsection{Zeeman splitting of the CB states in the QD}
A strong confinement of the electron and hole wave functions to the QD, as assumed by taking $V_c$ into account, corresponds to a splitting of the basis states into localized states which can be described as products of the former basis states and the confinement-induced envelope functions, 
\begin{equation}
\Psi^{j,\pm}_{\bm{m}}(\bm{r}) = \psi_{\bm{m}}(\bm{r})\ket{j,\pm}.
\end{equation}
We note that a non-trivial set of basis states requires $\max\{m_j\} \ge 3$.
We rewrite $H$ in a basis formed by the $\Psi^{j,\pm}_{\bm{m}}$ by taking the according matrix elements and find a new Hamiltonian $H_{d} $ describing the QD states.
We split $H_{d}$ into three parts,
\begin{equation}
H_{d} = H_{d}^{\mbox{\tiny d}} + H_{d}^{\mbox{\tiny bd}}+ H_{d}^{\mbox{\tiny bod}},\label{eq:QDHamiltonian}
\end{equation}
where $H_{d}^{\mbox{\tiny d}}$ denotes the diagonal elements of $H_{d}$, $H_{d}^{\mbox{\tiny bd}}$ denotes the block-diagonal parts of $H_{d}$ between the CB and VB, and $H_{d}^{\mbox{\tiny bod}}$ the associated block-off-diagonal elements. 
The external electric and magnetic fields are treated as a perturbation to the system. 
Hence diagonal terms of $H_{d}$ stemming from taking matrix elements of $H_Z, H_B,\mbox{ and } H_E$ are included in $H_{d}^{\mbox{\tiny bd}}$.
Since we are interested in describing electrons confined to CB states of the QD, we decouple the CB states from the VB states by a unitary transformation, the Schrieffer-Wolff transformation (SWT) $\tilde{H}_{d}=e^{-S} H_{d} e^{S}$, where $S$ is an anti-unitary operator  ($S^{\dagger} = -S$).\cite{winkler_spin-orbit_2003}
We approximate the SWT to third order in a small parameter $\lambda$ determined by the ratio of the CB-VB coupling and the CB-VB energy gap.
To this end, we express $S$ as  $S = S_1+S_2+S_3$, where $\mathcal{O}(S_i) = \lambda^{i}$. 
Here, the operators $S_i$ are defined by $[H_{d}^{\mbox{\tiny d}},S_1] = -H_{d}^{\mbox{\tiny bod}}$, $[ H_{d}^{\mbox{\tiny d}},S_2] =  -[H_{d}^{\mbox{\tiny bd}},S_1]$, $[H_{d}^{\mbox{\tiny d}},S_3] =  -[H_{d}^{\mbox{\tiny bd}},S_2]-1/3 [[H_{d}^{\mbox{\tiny bod}},S_1],S_1]$.\cite{winkler_spin-orbit_2003}
Since $\lambda$ is small, we can expand $e^S$ up to third order in $\lambda$ using the decomposition of $S$. 
Assuming that $\mathcal{O}(H_{d}^{\mbox{\tiny d}}) = \lambda^0$, $\mathcal{O}(H_{d}^{\mbox{\tiny bd}})= \mathcal{O}(H_{d}^{\mbox{\tiny bod}}) = \lambda^1$, we perform the SWT where we keep terms up to third order in $\lambda$ in the final Hamiltonian $\tilde{H}_{d}$.
In a last step, we project $\tilde{H}_{d}$ on the CB and find an effective CB Hamiltonian, $\tilde{H}_{d}^{\mbox{\tiny CB}}$. 
In $\tilde{H}_{d}^{\mbox{\tiny CB}}$, the single QD levels are strongly coupled, and thus cannot be treated perturbatively anymore.
Instead, we diagonalize  $\tilde{H}_{d}^{\mbox{\tiny CB}}$ exactly and evaluate the eigenenergies $E_{n}^{\pm}$, where the indices  denote the $n$th QD level from the VB edge with effective spin $\pm$. 
We find the $g$ factor of the $n$th spin-split QD level by taking
\begin{equation}
g_n=
\frac{
E_{n}^{+}-E_{n}^{-}
}{\mu_B |\bm{B}|},
\label{eq:def-g-factor}
\end{equation}
with Bohr magneton $\mu_B$. 
Since the exact values of the energies $E_{n}^{\pm}$ depend on the magnitude and direction of the external fields $\bm{E}$ and $\bm{B}$, $g_n = g_n(\bm{E}, \bm{B})$.
$\tilde{H}_{d}^{\mbox{\tiny CB}}$ contains higher order terms in $\bm{B}$; thus we find 
\begin{equation}
g_n = g_{n,0} + g_{n,2}  |\bm{B}|^2,
\end{equation}
which is consistent with the general  behavior expected of $\tilde{H}_{d}$ under time reversal.
However, with $|g_{n,2}|\ll |g_{n,0}|$, the quadratic dependence of $g_n$ on $ |\bm{B}|$ is barely measurable in experiments.

\begin{figure*}[t]
\centering
\includegraphics[width=.88\textwidth]{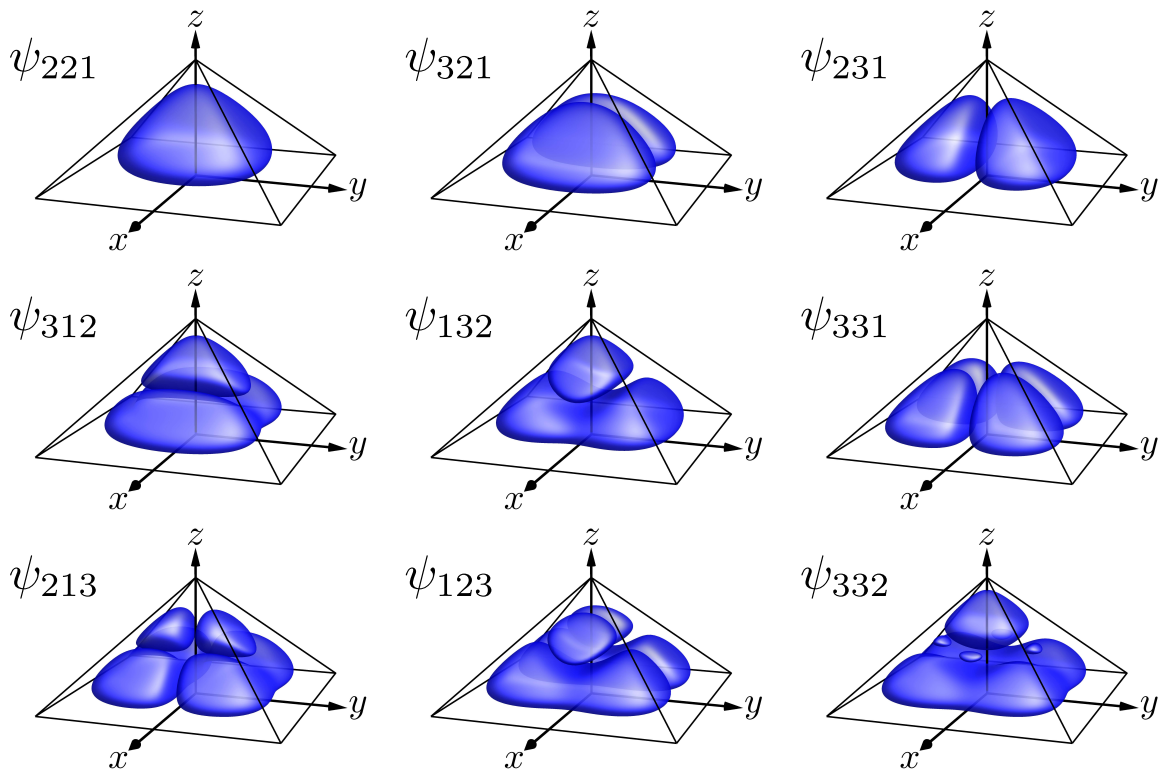}
\caption{
Probability distributions of the smallest nontrivial set of trial wave functions $\psi_{\bm{m}}(\bm{r})$, i.e.,\ $\max(m_i)=3$, satisfying the hard-wall boundary conditions for the geometry given in Fig.~\ref{fig:setup}. 
We show contour plots of $|\psi_{\bm{m}}(\bm{r})|^2=0.1$ inside the pyramidal geometry assumed for the QD; see Fig.~\ref{fig:setup}.
Note the degenerate pairs: 
$\psi_{321}$ and $\psi_{231}$,
$\psi_{312}$ and $\psi_{132}$,
$\psi_{213}$ and $\psi_{123}$.
}
\label{fig:particle-densities}
\end{figure*}

\begin{figure}[t]
\centering
\includegraphics[width=\columnwidth]{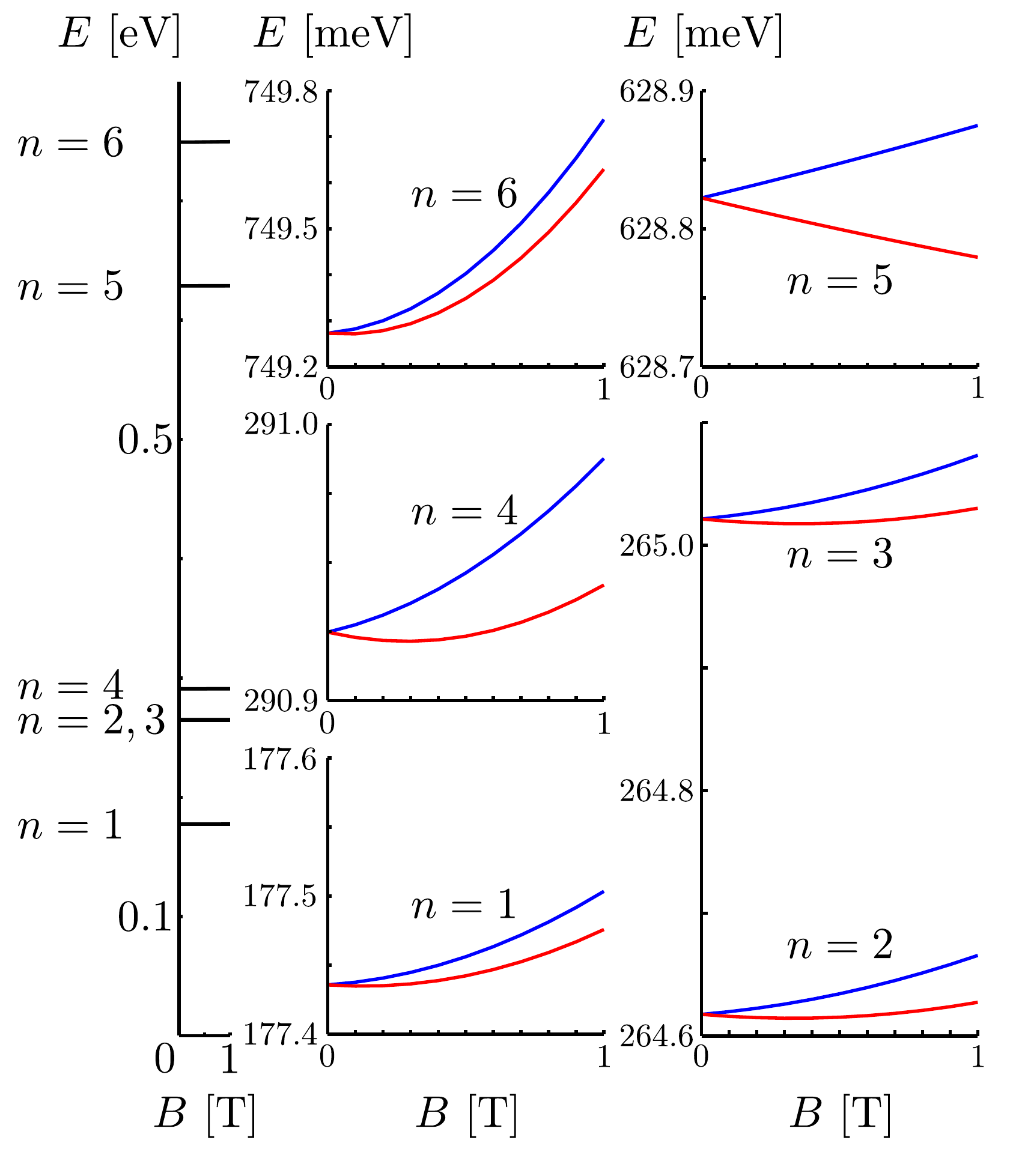}
\caption{
Left: Spectrum of the lowest six QD levels $E_n$ of $\tilde{H}_{d}^{\mbox{\tiny CB}}$ as a function of the magnetic field $\bm{B} = (0,0,B_z)$, where we increase $|\bm{B}| = 0 \mbox{ T to } 1 \mbox{ T}$. 
We assume $\bm{E}=0$.
Right: Enlarged plots of the $\bm{B}$-dependent splitting of the single QD levels. For most QD levels, except for $n=5$, we observe a nonlinear dependence of $E_{n}^{\pm}$ on $\bm{B}$.
\label{fig:spectrum}}
\end{figure}

\begin{figure}[t]
\centering
\includegraphics[width=\columnwidth]{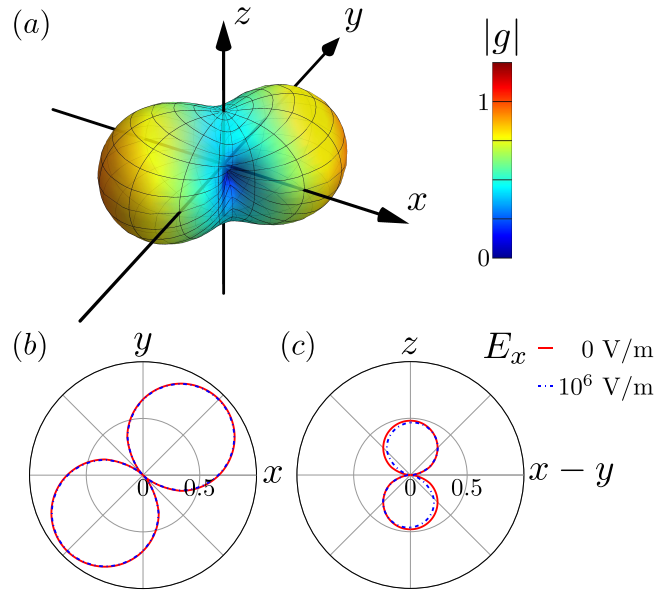}
\caption{
Ground state $g$ factor 
$|g_1|$
as a function of the magnetic field direction for $|\bm{B}| = 1 \mbox{ T}$ shown in
(a) 3D plot, and cuts along the planes (b) $xy$, and (c) $(x-y)z$ with electric field $\bm{E}=(E_x,0,0)$.}
\label{fig:gfplot-s1}
\end{figure}

\begin{figure}[!htb]
\centering
\includegraphics[width=\columnwidth]{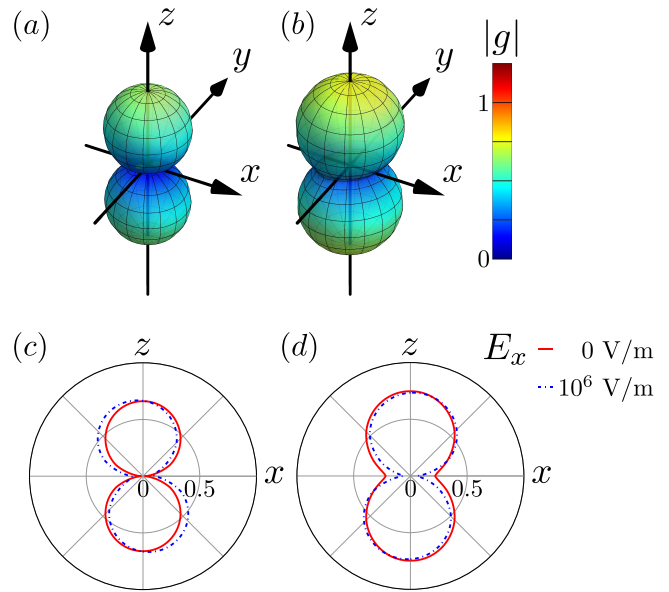}
\caption{$|g_2|$ and $|g_3|$ as functions of the magnetic field direction for $|\bm{B}| = 1 \mbox{ T}$ shown in 3D plots for
(a) $n=2$, (b) $n=3$, and cuts along the $xz$ plane  for (c) $n=2$ and (d) $n=3$ with electric field $\bm{E}=(E_x,0,0)$.}
\label{fig:gfplot-s2}
\end{figure}

\begin{figure}[!htb]
\centering
\includegraphics[width=\columnwidth]{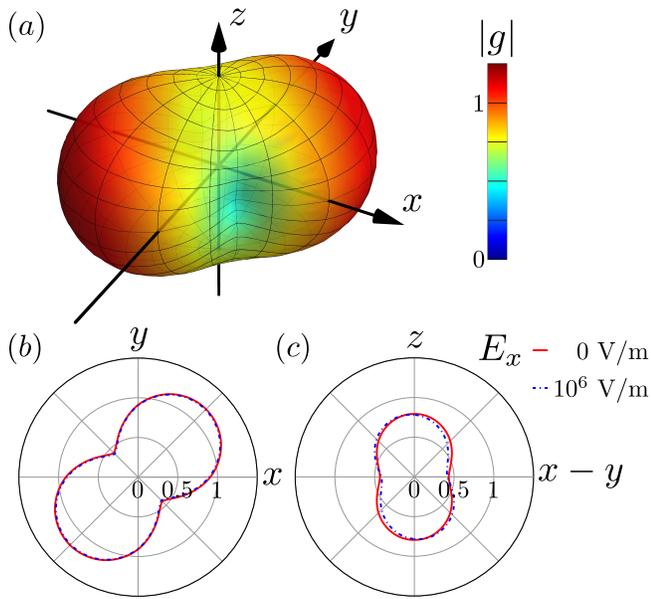}
\caption{$|g_4|$
as a function of the magnetic field direction for $|\bm{B}| = 1 \mbox{ T}$ shown in
(a) 3D plot, and cuts along the planes (b) $xy$, and (c) $(x-y)z$ with electric field $\bm{E}=(E_x,0,0)$.}
\label{fig:gfplot-s4}
\end{figure}

\begin{figure}[!htb]
\centering
\includegraphics[width=\columnwidth]{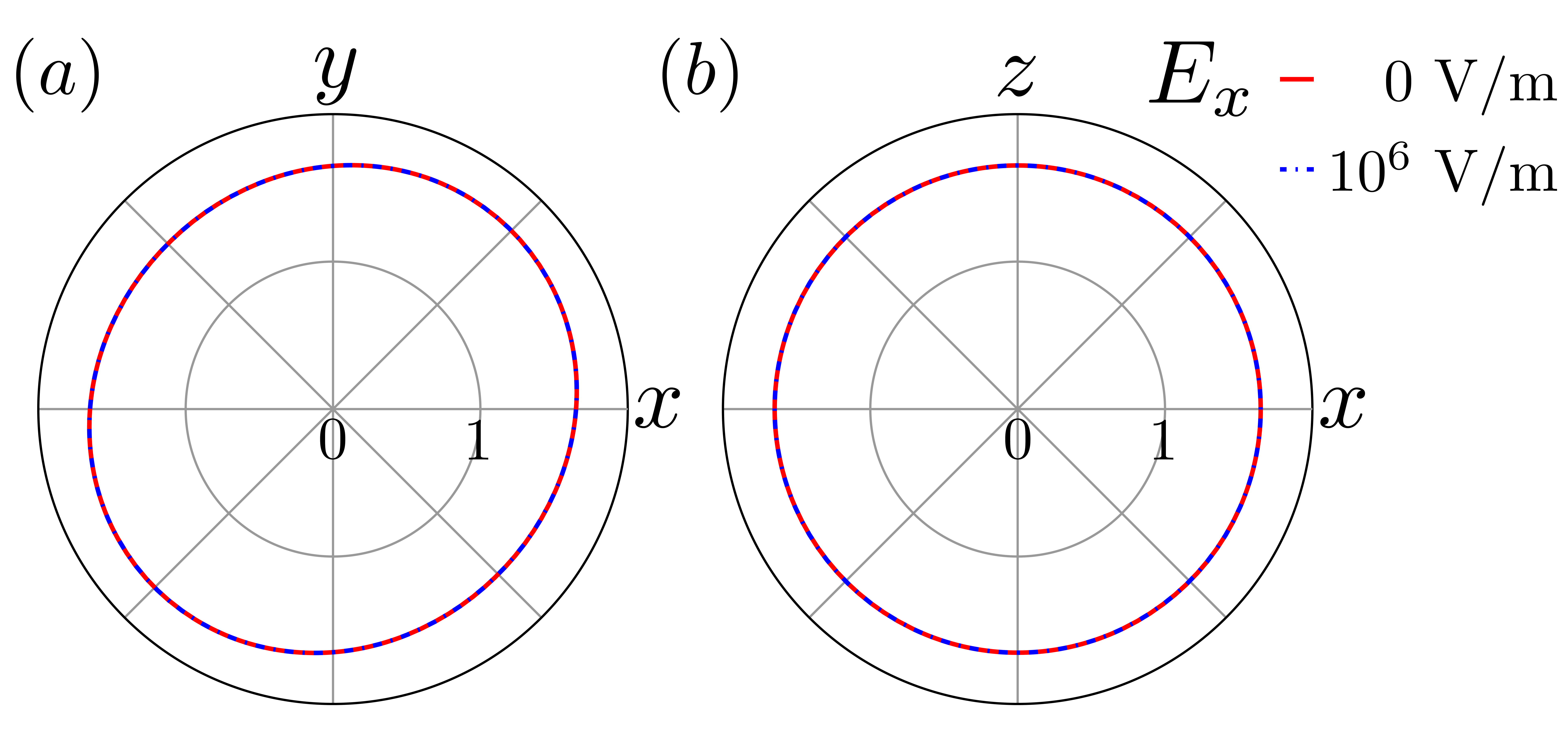}
\caption{$|g_5|$
as a function of the magnetic field direction shown  for $|\bm{B}| = 1 \mbox{ T}$ in cuts along the planes 
(a) $xy$, and (b) $xz$ with electric field $\bm{E}=(E_x,0,0)$.
Here we omit the 3D plot since the $g$ factor shows a spherical distribution, where such a plot does not yield further insight.}
\label{fig:gfplot-s5}
\end{figure}

\begin{figure}[!htb]
\centering
\includegraphics[width=\columnwidth]{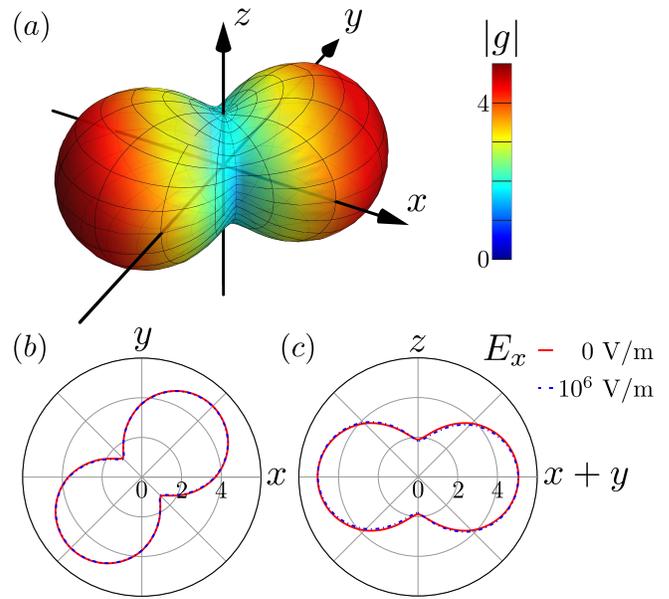}
\caption{$|g_6|$
as a function of the magnetic field direction  for $|\bm{B}| = 1 \mbox{ T}$ shown in
(a) 3D plot, and cuts along the planes (b) $xy$ and (c) $(x+y)z$ with electric field $\bm{E}=(E_x,0,0)$.
Note that the color scale changed because $|g_6|$ reaches larger values than the $|g_n|$ of the QD levels with $n<6$.}
\label{fig:gfplot-s6}
\end{figure}

\section{Results \label{sec:results}}
In this section, we present the results of the calculations outlined in Sec.~\ref{sec:model}. 
All calculations were performed for a pyramidal QD of height $a = 50\mbox{ nm}$. 
We consider basis states that fulfill $\max\{m_i\}\le 3$, which results in a splitting of each band $\ket{j,+}$ ($\ket{j,-}$) into nine QD levels.
The system parameters used for the Hamiltonians are listed in Table \ref{tab:constants} in Appendix~\ref{app:param}, where the notation directly corresponds to the notation used in Ref.~\onlinecite{winkler_spin-orbit_2003}.
\subsection{Probability distribution of the wave function\label{sec:res:prob}}
We show contour plots of the probability distribution $|\psi_{\bm{m}}(\bm{r})|^2$ of the wave function found in Eq.~(\ref{eq:3D-wave-function}), see Fig.~\ref{fig:particle-densities}.
We present the lowest-energy states forming the smallest nontrivial set of wave functions. 
The ground state $\psi_{221}$ with associated ground state energy $E_{221}=0.53\mbox{ meV}$ exhibits $s$-wave character; i.e.,\ we find a single density cloud roughly fitting the pyramidal shape.
For excited states, nodes appear in the center of the pyramid and along the axes of the coordinate system.
We observe $p$-wave character for the states $\psi_{321}$, $\psi_{231}$, $\psi_{312}$, and $\psi_{132}$; see Fig.~\ref{fig:particle-densities}.
The wave functions $\psi_{m_im_jm_k}$ and $\psi_{m_jm_im_k}$ with $m_i\ne m_j$ are degenerate and we find that the associated particle densities are of the same form, only with nodes oriented along different axes, i.e.,\ $x$ and $y$.
Further restrictions arising from the pyramid geometry, such as correlations between the coordinates, result in symmetries regarding the quantum numbers, $\psi_{m_im_im_j}=\psi_{m_jm_jm_i}$.

\subsection{Spectra of the CB states in the QD\label{sec:res:spec}}
In Fig.~\ref{fig:spectrum}, we plot the energy spectrum of the low-energy CB states given by $\tilde{H}_{d}^{\mbox{\tiny CB}}$ and examine the behavior of the QD levels as functions of  $\bm{B}= (0,0,B_z)$. 
For $|\bm{B}|=0$, we find six degenerate QD levels $E_n$ which split into pairs while increasing $\bm{B}$ from $0$ to $1\mbox{ T}$, where we assume that $\bm{E}=0$. 
Confinement and strain push the QD levels far apart from each other; hence the $\bm{B}$-induced spin splitting cannot be observed in the full plot, Fig.~\ref{fig:spectrum} on the left.
To circumvent this, we produce magnified plots showing the $\bm{B}$ dependence of the single QD levels $n$, Fig.~\ref{fig:spectrum} on the right.
We note that the splitting of the CB levels, $E_{n+1}-E_n$, is on the order of $100\mbox{ meV}$ which contrasts the Zeeman splitting, $E_n^{+}-E_n^{-}$, which is on the order of $1\mbox{ meV}$ or below.
For most QD levels $E_n^{\pm}$, we observe a clearly nonlinear dependence on $\bm{B}$, indicating a diamagnetic shift of the QD levels.\cite{van_bree_g_2012}
This dependence is not independent of the direction of $\bm{B}$, resulting in an anisotropy associated with the $g$ factor; see Sec.~\ref{sec:gfactorresults}.

\subsection{$g$ factor of the CB states in the QD\label{sec:gfactorresults}}
We discover strong anisotropies for the $g$ factors of electrons confined to low-energy CB states of pyramidal shaped InAs QDs.
The $g$ factors of the first six QD levels from the VB edge, $g_n$ with $n=1,\ldots,6$, are shown as 3D plots and cuts along specific planes in Figs.~\ref{fig:gfplot-s1} to \ref{fig:gfplot-s6} in ascending order.
We calculate the $g_n$ for magnetic fields of strength $|\bm{B}|=1\mbox{ T}$.
We further apply electric fields of strengths $|\bm{E}|=0\mbox{ V/m}$ and $|\bm{E}|=10^6\mbox{ V/m}$ along the $x$ axis.
In response to an electric field along the $x$ axis the anisotropy axis slightly tilts away from the $z$ axis.
To reduce calculation effort, we interpolate between data points, however, we have checked consistency in several cases with non-interpolated plots.

We find anisotropies of various shapes and directions depending on the QD level under consideration.
We observe the emerging of three main axes of anisotropy, $x+y$, $x-y$, and $z$, pointing along crystallographic directions $[110]$, $[1\bar{1}0]$, and $[001]$, respectively. 
QD levels $n=1,4,6$ ($n=2,3$) reveal $g$-factor maxima along the $x+y$ ($z$) axes, whereas small $g$-factor values tend to appear along(in) the $x-y$ axis ($xy$ plane).
Along the $x-y$ axis we observe that a special situation arises for $n=1$; here $g$ approaches a very small value close to but still larger than zero.
However, this drop depends strongly on the dot size; see Sec.~\ref{sec:gfactordiscussion}.
Interestingly, $g_5$ barely exhibits any anisotropy with maximum values at the $x+y$ axis and minimum values at the $x-y$ axis; see Fig.~\ref{fig:gfplot-s5}.
This is in contrast to $g_6$, where we note a considerable increase of the $g$-factor values and again a significant anisotropy.
Note the change of the color scale in Fig.~\ref{fig:gfplot-s6}.
In general, we observe a dependence of the absolute values of the $g_n$ on the QD size; see Sec.~\ref{sec:gfactordiscussion}.

\section{Discussion \label{sec:discussion}}
In this section, we comment on the probability distributions of pyramidal QDs calculated in Sec.~\ref{sec:res:prob}.
Furthermore, we discuss the $\bm{B}$ dependence of both spectrum and $g$ factor of the CB states in the QD presented in Secs.~\ref{sec:res:spec} and \ref{sec:gfactorresults}, respectively.
\subsection{Probability distribution of the wave function\label{sec:disc:prob}}
The wave functions of the lowest states exhibit the structure of cuboidal wave functions adapted to the pyramidal shape of the enclosing QD.
We definitely observe the ground state as well as excited states.
This is consistent with the method used for the construction of the wave functions.
Note that the wave functions $\psi_{\bm{m}}(\bm{r})$ are not exact eigenfunctions of the Schr\"odinger equation.
However, the boundary conditions are satisfied and the corresponding energies, see Eq.~(\ref{eq:HpsiEpsi}), are smaller than the energies of known analytical solutions of the Schr\"odinger equation provided that the correct boundary conditions are taken into account.\cite{horley_particle_2012}
Due to the method of construction, we find that the wave functions do not vanish at the diagonal planes $(x+y)z$ and $(x-y)z$, respectively, as was observed in Ref.~\onlinecite{horley_particle_2012}.
Furthermore, the authors of the work presented in Ref.~\onlinecite{horley_particle_2012} explicitly state that the obtained set of wave functions is incomplete; solutions with a finite density at the center of the pyramid are not contained. 
In particular, a distinct ground state is missing. 
From this we conclude that our set of wave functions is more suitable to describe low-energy states in pyramidal QDs.
Numerical calculations of QD wave functions usually include piezoelectric potentials and specific material properties directly from the beginning, which complicates a direct comparison.\cite{grundmann_inas/gaas_1995,stier_electronic_1999}
However, compared to numerical calculations without strain as performed in Ref.~\onlinecite{grundmann_inas/gaas_1995}, where the wave functions extend into a wetting layer, and Ref.~\onlinecite{stier_electronic_1999}, where no intermixing with a wetting layer is observed, we report similar shapes of the probability distributions with our analytical ansatz.
Even though we apply this simplistic model, we recover the effects recently observed in experiments to a very good degree;\cite{takahashi_electrically_2013} see Sec.~\ref{sec:gfactordiscussion}.

\subsection{Spectra of the CB states in the QD}
After diagonalizing $\tilde{H}_{d}^{\mbox{\tiny CB}}$, we find states in the CB of the QD which are degenerate for $|\bm{B}|=0$ and split into pairs by an increasing magnetic field.
These energy levels exhibit a quadratic dependence on $\bm{B}$.
We note that the direction of the magnetic field is important to the exact behavior of the splitting of the QD levels.
Due to the highly admixed nature of the final eigenstates of $\tilde{H}_{d}^{\mbox{\tiny CB}}$, which consist of CB and VB states of the basis introduced for $H_d$ in Eq.~(\ref{eq:QDHamiltonian}), we find ourselves unable to comment on the exact shape of the $n$th eigenfunction.
For illustrative plots of the electron wave function in considerably (one order of magnitude) smaller QDs from numerical and experimental studies, we refer the interested reader to Refs.\ \onlinecite{stier_electronic_1999} and \onlinecite{vdovin_imaging_2000}.

\subsection{$g$ factor of the CB states in the QD \label{sec:gfactordiscussion}}
The reported anisotropy in our system stems from several effects.
The first effect is the mixing of CB and VB states caused by the confinement potential and intrinsic material parameters of the QD.
This mixing is further influenced by the second effect, a change of gaps between the bands $\ket{j,\pm}$ due to strain.
The intrinsic strain fields in the QD impose additional constraints on the system yielding a reduction of the symmetry of the level splitting with respect to the direction of $\bm{B}$.
Furthermore, the strain fields reduce the symmetry class of the pyramid along the $z$ axis from $C_4$ to $C_2$.\cite{grundmann_inas/gaas_1995}
This reduction of the symmetry class agrees well with the observed anisotropy of the $g_n$ in our work.
Additionally, effects due to the orbital coupling of $\bm{B}$ may have an effect on $g$.
For $|\bm{B}| = 1\mbox{ T}$, we find that the magnetic length $l_B=\sqrt{\hbar/e |\bm{B}|}\sim 25\mbox{ nm}$ is much smaller than the dot size characterized by $a=50\mbox{ nm}$; hence Landau levels form.
However, we took this into account by including $H_B$ into our Hamiltonian; see Eq.~(\ref{eq:hamiltonian}).
Compared to experimental results,\cite{takahashi_electrically_2013} we observe very small $g$ factors, mainly $g_n<2$. 
However, small $g$ factor values, in particular a zero crossing of $g$ due to the transition from the bulk value $g_{\text{bulk}}\approx-14.9$ to the free electron value $g_{\text{free}} = +2$, have also been reported for circular and elliptical InAs QDs.\cite{pryor_lande_2006,pryor_erratum:_2007}
This transition is characterized as a function of the band gap between the CB and VB in the QD. 
In fact, we find a comparable magnitude of the $g$-factor values considering the band gap present in our system.
In general, decreasing the QD size leads to a decrease of the CB-VB admixture and the $g$-factor values ultimately yield the free electron value, $g_{\text{free}} = +2$.
On the other hand, when increasing the QD size the $g$-factor values will finally approach the bulk value, $g_{\text{bulk}}\approx-14.9$.
Considering these two limits and assuming that the $g$ factor is a continuous quantity, zero values of $g$ will be observed eventually.\cite{pryor_lande_2006,pryor_erratum:_2007}\\

\section{Comparison to experiment \label{sec:comparison}}
In this section, we compare our results to recent experimental observations of the three-dimensional $g$-factor anisotropy in self-assembled InAs QDs by Takahashi {\textit{ et al.}}; see Ref.~\onlinecite{takahashi_electrically_2013}. 
The anisotropy of the QD $g$ factor is usually extracted by transport measurements for different magnetic field directions.\cite{takahashi_electrically_2013,dhollosy_g-factor_2013}
The basic setup of these experiments consists of a QD which is tunnel coupled to two leads.
An additional back-gate creates an electric field parallel to the growth direction.
The back-gate voltage is used to select the QD level participating in the transport by changing the chemical potential of the QD.
Furthermore, the tunneling rates depend on the different $g$ factors of QD and leads.\cite{stano_spin-dependent_2010}
We first point out that the  QD considered in Ref.~\onlinecite{takahashi_electrically_2013} is rather a half pyramid due to the applied gates.
Thus, deviations of the absolute value of $g$ compared to our findings are not unexpected. Such deviations increase even further due to different dot sizes.
However, we find good qualitative agreement when accounting for the different confinement geometries in the following way.
One can perform a coordinate transformation in order to align the axes of the upright pyramid considered above and the half-pyramid of Ref.~\onlinecite{takahashi_electrically_2013}.
Indeed, a rotation of  $~135^\circ$ around the $y$ axis aligns the symmetry axes of both systems in first approximation.
We observe now  that the $g$-factor anisotropies of the QD levels $n=2,3$ (Fig.~\ref{fig:gfplot-s2}) agree well with regions I and II of the charge stability diagram reported by Takahashi {\textit{ et al.}} in Ref.~\onlinecite{takahashi_electrically_2013}.
In region III they also find a state with a spherical distribution of the $g$ factor similar to our calculation for QD level $n=5$.
Furthermore, they report measurements of a symmetrically covered upright pyramid as well.
In this case the axes and shapes of the anisotropy are directly comparable to our results.
The associated $g$-factor anisotropy agrees well with our findings for QD levels $n=2,3$.
In general, due to confinement and strain, the QD size and shape have a strong influence on characteristic quantities such as spectrum and $g$ factor, both absolute value and anisotropy.
However, we find good qualitative agreement between our model calculation and the measurements.
This is not surprising since both consider square-based pyramids which conserve the main anisotropy axes independent of the QD size.
Finally, we point out that our model further predicts different shapes of the $g$-factor anisotropy depending on the QD levels --
in particular, shapes not yet observed in experiments, such as the ones described for the QD levels $n=1,4,6$.

\section{Conclusion \label{sec:conclusion}}
In conclusion, we have found trial wave functions satisfying hard-wall boundary conditions for a pyramidal QD geometry. 
We calculated the associated particle density distributions of the low-energy states and found a ground-state-like, $s$ symmetric state of lowest energy, as well as excited states with nodes along the coordinate axes of the system and at the center of the QD. 
We argued that these wave functions provide a good basis for analytical calculations of QD states.
Furthermore, we have presented 8-band calculations to derive the spectrum of low-energy CB states in the QD.
The magnetic field induced splitting of the QD levels shows a nonlinear dependence on the magnetic field and strong anisotropies depending on the direction of the field. 
Starting from this, we have calculated the $g$ factor of low-energy electrons in self-assembled InAs QDs subject to externally applied electric and magnetic fields.
We calculated the $g$ factor for all possible spatial orientations of the magnetic field and found distinct anisotropies. 
In particular, we showed that the anisotropies include configurations where the $g$ factor drops down to values close to zero.
Furthermore, we observed that the shape of the anisotropies depends on the QD level $n$ and that the maximal values of $g_n$ increase with $n$.
Finally, we showed that our results are in good qualitative agreement with recent measurements.
From these findings we conclude that the direction of magnetic fields applied to QDs can be used to control the splitting of qubit states efficiently and hence 
should prove useful for the manipulation of qubits in such QDs.

\section{Acknowledgments}
We thank Christoph Kloeffel, 
Markus Samadashvili, 
Peter Stano, Dimitrije Stepanenko, and Seigo Tarucha for useful discussions. 
We acknowledge support from the Swiss NF, the NCCR QSIT, and IARPA.

\begin{appendix}
\section{Trial wave functions\label{app:wave-function}}
The Schr\"odinger equation of a particle confined to a square with sides of length $a$,
\begin{equation}
-\frac{\hbar^2}{2m_0}
\left(
\frac{d^2}{dx^2}+\frac{d^2}{dy^2}
\right)
\psi^{\Box}(x,y)
=E^{\Box}\
\psi^{\Box}(x,y),
\end{equation}
with boundary conditions
$\psi^{\Box}(x,y)=0$ for $x=0$, $y=0$, $x=a$ or $y=a$,
has the well-known solution:
\begin{align}
\label{eq:wave function}
\psi_{mn}^{\Box}(x,y)=&\
\frac2a\
\sin\left(\frac{m\pi}ax\right)\
\sin\left(\frac{n\pi}ay\right),
\\
\label{eq:energies}
E_{mn}^{\Box}=&\
\frac{\hbar^2\pi^2}{2m_0a^2}\
\left(m^2+n^2\right).
\end{align}
The wave function of a particle confined in an isosceles triangle obtained by cutting the square along the diagonal, $\psi^{\triangle}(x,y)$, is constructed by symmetric and asymmetric
linear combination of degenerate solutions to the square problem, $\psi_{mn}^{\Box}$ and $\psi_{nm}^{\Box}$,\cite{li_particle_1984} and we find
\begin{align}
\psi_{mn}^{\triangle s}=&
\tfrac{1}{\sqrt2}
\Big(\psi_{mn}^{\Box}+\psi_{nm}^{\Box}\Big),
\\
\psi_{mn}^{\triangle a}=&
\tfrac{1}{\sqrt2}
\Big(\psi_{mn}^{\Box}-\psi_{nm}^{\Box}\Big),
\end{align}
where $\psi_{mn}^{\triangle s}$ ($\psi_{mn}^{\triangle a}$) vanishes at $x+y=a$ for $m+n$ odd (even). 
The general wave function takes the form
\begin{equation}
\psi_{mn}^{\triangle}=
\frac{1}{\sqrt2}
\Big(\psi_{mn}^{\Box}+(-1)^{m+n+1}\psi_{nm}^{\Box}\Big),
\end{equation}
with $m,n=1,2,3,\dots$ and $m\ne n$ to prevent the construction of a vanishing wave function $\psi_{mm}^{\triangle}=0$.
We apply a coordinate transformation characterized by
$x=-[\tilde x+(\tilde y-\tilde a)]/2\sqrt{2}$ and $y=[\tilde x-(\tilde y-\tilde a)]/2\sqrt{2}$ 
in order to bring the triangle into upright position, i.e.,\ the apex of the triangle is centered above the base, and find
\begin{equation}
\psi_{mn}^{\triangle}(\tilde x,\tilde y)=
-\psi_{mn}^{\triangle}\Big(\tfrac{\tilde x+(\tilde y-\tilde a)}{2\sqrt{2}},\tfrac{\tilde x-(\tilde y-\tilde a)}{2\sqrt{2}}\Big)
\end{equation}
with $m,n=1,2,3,\dots$, $m\ne n$, and $\tilde a=a/\sqrt{2}$.

\begin{figure}[!tb]
\includegraphics[width=.2775\textwidth]{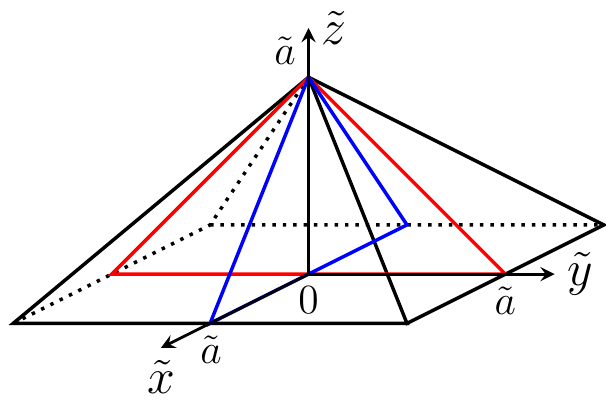}
\caption{We span the pyramid volume by multiplying two upright isosceles triangles (red and blue).
Note that $\tilde{\bm r}$ and $\tilde a$ correspond to $\bm{r}$ and $a$ in the main text, respectively.}
\label{fig:span-pyramid}
\end{figure}

Starting from the solution to the two-dimensional Schr\"odinger equation, we construct an ansatz or trial wave function that is not an eigenfunction of the three-dimensional (3D) Schr\"odinger equation but nonetheless fulfills the boundary conditions of the pyramid and expected symmetries.
We span the 3D volume of the pyramid with the product of two upright triangles, see Fig.~\ref{fig:span-pyramid}, and find the wave function
\begin{equation}
\begin{split}
\psi_{\bm{m}}(\tilde{\bm{r}})=&
\ c \
\psi_{m_{x}m_{z}}^{\triangle}(\tilde x,\tilde z)\
\psi_{m_{y}m_{z}}^{\triangle}(\tilde y,\tilde z)\\
=&
\ c\prod_{\xi=\tilde x,\tilde y}
\Big[
\sin\left(\alpha_{\xi}\ \xi^{+}\right)
\sin\left(\alpha_{z}\ \xi^{-}\right)
\\&
-(-1)^{m_{\xi}+m_{z}}
\sin\left(\alpha_{z}\ \xi^{+}\right)
\sin\left(\alpha_{\xi}\ \xi^{-}\right)
\Big],
\end{split}
\end{equation}
with
$\tilde{\bm{r}}=(\tilde x,\tilde y,\tilde z)$,
$c=\csc(\pi \tilde z)/N_{\bm{m}}$, 
$\alpha_{i}=m_{i}\pi/\tilde a$, $m_{i}=1,2,3,\dots$, $m_{x}\ne m_{z}$, $m_{y}\ne m_{z}$, $\bm{m} = (m_x, m_y, m_z)$, 
$\xi^{\pm}=\xi\pm(\tilde z-\tilde a)/2$, and $N_{\bm{m}}$ such that the integral over the pyramid volume yields $\int d^3\tilde r\ |\psi_{\bm{m}}(\tilde{\bm{r}})|^2\equiv 1$.
Note that we have added the term  $\csc(\pi \tilde z)$ in order to restore the asymptotes at the apex and the base to the correct power law behavior in $z$ that were altered by taking the product $\psi_{m_{x}m_{z}}^{\triangle}\psi_{m_{y}m_{z}}^{\triangle}$.
This factor is essential for obtaining $s$- and $p$-wave like states.
The energies of state $\psi_{\bm{m}}$ are given by
\begin{equation}
E_{\bm{m}}=\frac{\hbar^2}{2 m_0}\bra{\psi_{\bm{m}}(\tilde{\bm{r}})}(-i \nabla)^2 \ket{\psi_{\bm{m}}(\tilde{\bm{r}})}.
\end{equation}
For notational simplicity we use $\psi_{\bm{m}}\equiv\psi_{m_xm_ym_z}$. 
We note that the states $\psi_{m_xm_xm_z}$ and $\psi_{m_zm_zm_x}$ coincide by construction and that $\psi_{m_xm_ym_z}$ and $\psi_{m_ym_xm_z}$ are degenerate.

As mentioned above, $\psi_{\bm{m}}$ is not an eigenfunction of the 3D Schr\"odinger equation.
However, the boundary conditions are fulfilled.
In addition, the energies $E_{\bm{m}}$ are smaller than the eigenenergies of known analytical solutions provided that the correct boundary conditions at the base of the pyramid are taken into account.\cite{horley_particle_2012}
Furthermore, the set of eigenfunctions reported in Ref.~\onlinecite{horley_particle_2012} is incomplete and in particular lacks the ground state and states with a non-vanishing particle density (of $s$-wave type) at the center of the pyramid.
In contrast, our trial wave functions form a complete set including states with $s$- and $p$-wave character. 
Despite the fact that $\psi_{\bm{m}}$ is not an eigenfunction, we conclude that our trial wave functions provide a good starting point for analytical investigations of pyramidal quantum dots.

\FloatBarrier
\section{Material parameters\label{app:param}}
We choose the notation for the parameters exactly as given in Ref.~\onlinecite{winkler_spin-orbit_2003}. See Table \ref{tab:constants}.
\FloatBarrier
\begin{table}[!h]
\begin{tabular}{lldclld}
\hhline{===~===}
$E_g$ &[eV]& 0.418 &\phantom{xx}
&
$q$ && 0.39 
\\
$\Delta_0$ &[eV]& 0.380 &
&
$C_1$ &[eV]& -5.08\cite{winkler_spin-orbit_2003,vurgaftman_band_2001}
\\
$P$ &[eV\AA]& 9.197 &
&
$D_d$ &[eV]& 1.\cite{winkler_spin-orbit_2003,vurgaftman_band_2001}
\\
$C_k$ &[eV\AA]& -0.0112 &
&
$D_u$ &[eV]& 2.7
\\
$m^*$ &[$m_0$]& 0.0229 &
&
$D_u'$ &[eV]& 3.18
\\
$g^*$ && -14.9 &
&
$C_2$ &[eV]& 1.8\cite{bir_effect_1962,vurgaftman_band_2001}
\\
$\gamma_1$ && 20.40 &
&
$D'$ && -2.\cite{[{Due to lack of experimentally validated InAs parameters, we use InSb values which are assumed to be close to InAs values. }] [{}]trebin_electrons_1988,*weiler_warping-_1978}
\\
$\gamma_2$ && 8.30 &
&
$C_4$ &[eV\AA]& 11.3\cite{ranvaud_quantum_1979,silver_strain-induced_1992}
\\
$\gamma_3$ && 9.10 &
&
$C_5$ &[eV\AA]& 103.3\cite{ranvaud_quantum_1979,silver_strain-induced_1992}
\\
$B_{8v}^+$ &[eV\AA$^2$]& -3.393&
&
$C_5'$ &[eV\AA]& 76.9\cite{trebin_quantum_1979}
\\
$B_{8v}^-$ &[eV\AA$^2$]& -0.09511&
&
$a_{\text{\tiny InAs}}$ &[nm]&6.0583
\\
$B_{7v}$ &[eV\AA$^2$]& -3.178&
&
$a_{\text{\tiny GaAs}}$ &[nm]&5.65325
\\
$\kappa$ && 7.60 &
&
$\nu_{\text{\tiny InAs}}$ && 0.35\cite{levinshtein_handbook_1996}\\
\hhline{===~===}
\end{tabular}
\caption{
Material parameters used in this work.
If not stated otherwise, the parameters were taken from Ref.~\onlinecite{winkler_spin-orbit_2003}.
}
\label{tab:constants}
\end{table}

\end{appendix}
\newpage
\bibliography{lib}

\end{document}